# Dendrites and Efficiency: Optimizing Performance and Resource Utilization

**Roman Makarov[1,2], Michalis Pagkalos[1,2] and Panayiota Poirazi[1]**

1 – Institute of Molecular Biology and Biotechnology (IMBB), Foundation for Research and Technology Hellas (FORTH), Heraklion, 70013, Greece

2 – Department of Biology, University of Crete, Heraklion, 70013, Greece

Corresponding author: Panayiota Poirazi (poirazi@imbb.forth.gr)

## Highlights

- Dendrites optimize information processing and storage under resource constraints.
- Compartmentalization in dendrites is a key to enhance computational efficiency.
- Dendrites facilitate predictive, context-dependent processing in hierarchical networks.
- Dendritic nonlinearities might have evolved to process natural stimuli on behavioral timescales.
- Dendrites increase memory capacity and help balance memory binding and interference.

## Abstract

The brain is a highly efficient system evolved to achieve high performance with limited resources. We propose that dendrites make information processing and storage in the brain more efficient through the segregation of inputs and their conditional integration via nonlinear events, the compartmentalization of activity and plasticity and the binding of information through synapse clustering. In real-world scenarios with limited energy and space, dendrites help biological networks process natural stimuli on behavioral timescales, perform the inference process on those stimuli in a context-specific manner, and store the information in overlapping populations of neurons. A global picture starts to emerge, in which dendrites help the brain achieve efficiency through a combination of optimization strategies balancing the tradeoff between performance and resource utilization.

# Introduction

For animals, survival in the wild depends heavily on their ability to perform well on behavioral tasks such as recognizing predators, interpreting social cues, and remembering safe routes to food and shelter. Therefore, their sensory systems must be optimized for processing natural stimuli on behavioral timescales. To achieve this, sensory information must be accurately represented in the form of neuronal activity and then integrated and interpreted in a contextually appropriate manner. Furthermore, important information must also be efficiently stored and retrieved for future use. However, these processes come at a cost. Neuronal communication is metabolically expensive and the capacity for memory storage in the brain is limited. As a result, brains have evolved a variety of strategies, such as predictive coding [1] and sparse firing [2,3], to process and store information more efficiently. Growing evidence suggests that these strategies heavily rely on dendrites, the thin processes that extend from the cell bodies of neurons, to balance the tradeoff between high performance and resource savings **(Figure 1)**.

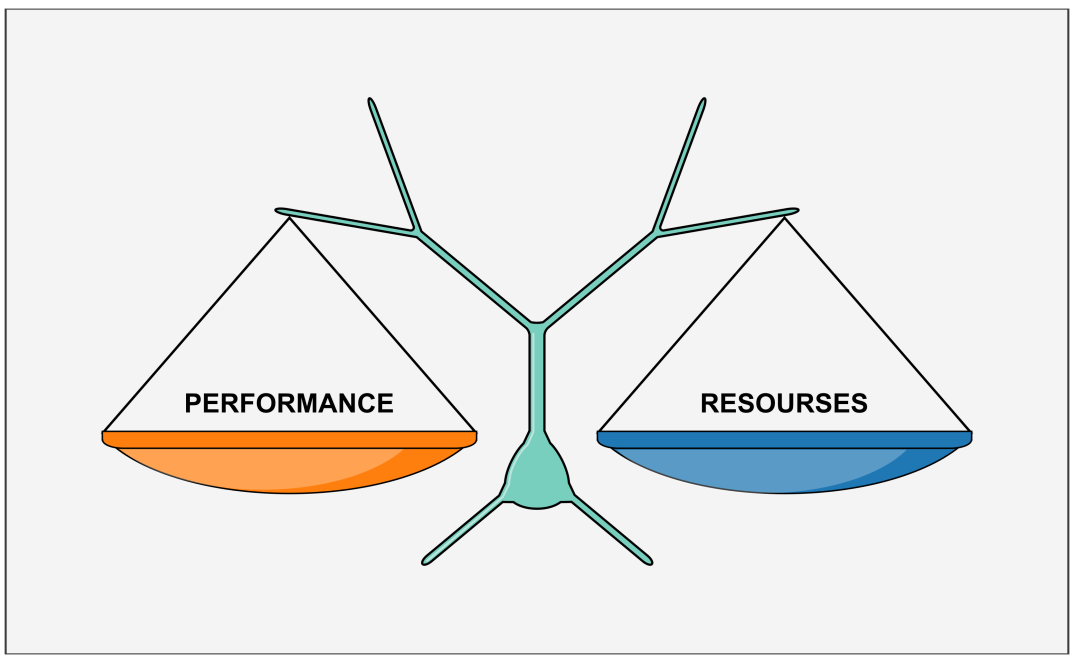

**Figure 1 |** Under evolutionary pressure, neurons in the brain must find a balance between high performance and resource savings. On one plate of this conceptual pair of scales, there is a demand to process, interpret and store information important for behavioral tasks. On the other plate, there are constraints on resources, such as energy and space. Here we propose that dendrites could help balance this tradeoff through various strategies, including hierarchical predictive coding, optimizing computations for natural stimuli, increasing expressivity, mitigating noise, and optimizing learning and storage capacity.

Dendrites of most excitatory neuron types receive thousands of synaptic inputs, distributed across their bifurcating branches. This morphological organization results in electrical and biochemical compartmentalization within a neuron. Thanks to a diverse reservoir of ionic and synaptic conductances, semi-independent dendritic compartments exhibit local regenerative events called dendritic spikes, that amplify postsynaptic potentials and mediate local plasticity. Dendritic events can either be isolated, promoting the formation of parallel computing units, or interact with each other while propagating to the soma, resulting in global activity. Together, the segregation, amplification, and nonlinear integration of activity within dendrites give rise to a diverse range of input-output transformations that a neuron can perform. Due to this computational complexity, dendrites are proposed to optimize neural processing in hierarchical networks, mitigate variability (noise) inherent to biological systems and enable neurons to process natural stimuli on behavioral timescales.

In this article, we review the latest findings on the role of dendrites in the efficient processing and storage of information. First, we discuss how dendrites underlie the segregation of inputs and the compartmentalization of activity in principal neurons. Then, we discuss the role of dendrites in specific optimization strategies, such as hierarchical predictive coding, optimizing computations for natural stimuli, increasing expressivity, mitigating noise, and increasing storage capacity. For each of these strategies we highlight how dendrites help to improve computational performance and save valuable resources.

## Compartmentalization as a cornerstone of dendritic optimization strategies

Due to their anatomical and biophysical characteristics, dendritic trees exhibit electrical and biochemical compartmentalization which give rise to local computations and plasticity. As a result, dendritic compartments can act as semi-independent thresholded units that nonlinearly integrate synaptic inputs before sending them to the soma [4–6] and have been suggested to serve as functional units of computation in the brain [7]. Payer et al. (2019) [8] delineate four classes of complex dendritic processing, among which information selection (when only a small portion of inputs is chosen for propagation) and routing (when the relative potency of dendritic subunits is modulated) heavily rely on compartmentalization. In this article we emphasize the importance of dendritic compartmentalization for efficient processing and storage of information in neuronal networks. We first show how compartmentalization is implemented in dendrites and then discuss particular optimization strategies based on compartmentalization in pyramidal neurons.

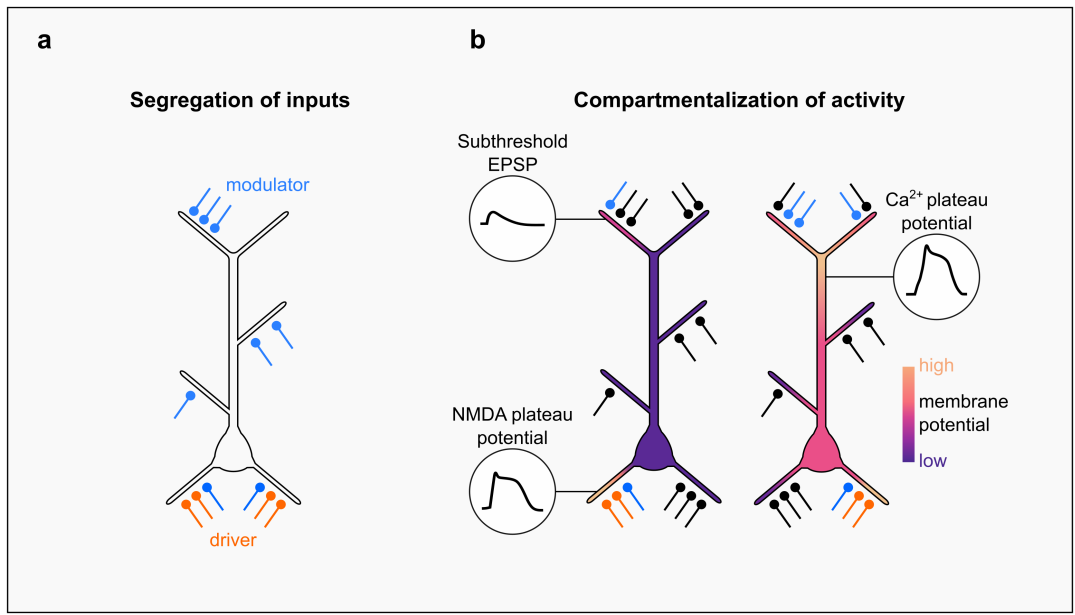

**Figure 2 | Compartmentalization in pyramidal neurons**

**(a)** Morphological structure of dendritic trees in pyramidal neurons enables distribution of synaptic inputs across different branches and functional separation of input signals on drivers (blue) and modulators (orange) of neuronal activity.

**(b)** Synaptic segregation and morpho-electric properties of dendrites result in subcellular compartmentalization of electrical activity. Dendritic activity can be locally restricted within specific dendritic branches in the form of postsynaptic potentials (PSPs) and NMDA spikes (left) or globally integrated, involving multiple compartments of a single neuron, as in the case of backpropagation-activated calcium spikes (BACs; right).

---

The elaborate, tree-like structure of dendrites has two important implications. First, it enables the distribution of synaptic inputs among different branches, where they can be processed in different ways. For example, spatial segregation in cortical pyramidal neurons allows for the functional separation of input signals that serve as driver vs. modulatory ones, carrying sensory and contextual information, respectively (**Figure 2a**). In cortical regions, feedforward (driver) inputs typically target the basal dendrites of pyramidal neurons while feedback (modulatory) inputs from higher areas are more commonly found in apical dendrites [9]. The coincidence of feedforward and feedback inputs is thought to associate different information streams [10] and determine the selectivity of neuronal responses to important stimuli [11]. Besides feedback connections, the modulatory input can be also represented by horizontal connections [12], thalamic inputs [13], inhibition [14–16] and neuromodulation [17].

Second, the tree-like structure of dendrites shapes the propagation of electrical signals through the cell, leading to subcellular compartmentalization of electrical activity. Postsynaptic signals enter a complex maze of cable-like branches with multiple bifurcation points, where they attenuate and often end up confined within a single branch [18]. However, active voltage-dependent conductances, which are abundant in dendrites, can help to overcome this phenomenon. Large depolarizations can evoke local regenerative events, such as dendritic Na+ spikes, and long-lasting NMDA and Ca2+ plateau potentials [19]. While some of these events are restricted within specific dendritic branches, others can be global, involving large parts of the dendritic tree (for a review see Stuyt et al., 2022 [20]; **Figure 2b**).

The exact degree of dendritic electrical compartmentalization remains a subject of debate. Early in vitro and computational studies suggested a high degree of compartmentalization while recent in vivo experiments claim that isolated dendritic events are much rarer than expected (for a review see Francioni et al., 2022 [21]). These new observations suggest an important role of somato-dendritic coupling, in line with the earlier studies on the segregation of inputs, portraying neurons as being organized into a few functional domains. In this paradigm, global somato-dendritic events, such as backpropagation-activated calcium spikes (BACs) serve as the mechanism for integrating modulatory and driver signals at the cellular level [20]. Somato-dendritic coupling has been suggested to play a major role in sensory detection [11], and explain conscious processing [22] and the effects of anesthesia [23]. Notably, through computational modeling Wybo et al. (2019) [24] demonstrated that pyramidal neurons comprise substantially fewer functional subunits than dendritic branches. Moreover, their results suggest that compartmentalization can be dynamically regulated. For example, topology (subunit extent and number) can be modified by balanced inputs or shunting inhibition in a context-dependent manner.

Overall, dendritic compartmentalization provides a powerful mechanism for flexibly modulating the neuronal output in response to spatio-temporally organized synaptic input. The ways in which compartmentalization underlies efficient neuronal processing are detailed in the following paragraphs.

# Information processing

### Optimizing hierarchical predictive coding

Hierarchical coding refers to the organization of neural processing in a hierarchy, from low-level sensory features to high-level abstractions. In the visual cortex, for example, primary areas process simple edges and lines, while higher areas process more complex shapes and scenes. In this way, features common for multiple visual objects can be reused across the hierarchy, reducing the need for redundant processing. Moreover, hierarchical coding allows for the top-down control of sensory processing and nonlinear mixing of information from different sensory modalities to create a more complete representation of the environment. It is believed that one of the primary computations performed by the cortex is an inference process, where bottom-up (sensory) information is combined with top-down expectations from prior knowledge to find a consistent explanation of sensory data [25]. Therefore, hierarchical coding is crucial for the brain to efficiently process and interpret sensory information, and to generate flexible and adaptive behaviors.

Recent studies suggest that dendrites can offer a mechanistic explanation of how hierarchical predictive coding (hPC) is implemented in the brain. Classical hierarchical predictive coding theory relies on error computations within each layer. However, Mikulasch et al. (2023) [25] have proposed a novel approach by shifting error computation from a separate neural population to the dendritic compartments of layer 2/3 pyramidal neurons (**Figure 3a**). This allows error computation to be performed in the voltage dynamics of individual dendritic compartments. According to this theory, a cortical neuron receives input from the lower level in hierarchy on its basal dendrites. Given enough "prior knowledge" the neuron can predict feedforward input and cancel it through lateral inhibition by parvalbumin-positive (PV) interneurons. On the other hand, novel unexpected inputs cannot be balanced by inhibition, so the signal reaches the soma. Thus, unexpected input results in the bottom-up prediction error. At the same time, the neuron receives predictions of its own activity from higher areas impinging on its apical dendrite. A mismatch between apical prediction signal and somatic spiking results in top-down prediction error [26]. Overall, the dendritic hPC theory highlights that by assigning the inference process to dendrites, rather than neuronal populations, the brain can save a significant amount of resources that can be allocated for other tasks.

## Processing natural stimuli on behavioral time scales

Beside feedback connections from the higher-order areas, a substantial fraction of horizontal connections also carry contextual information. For example, in the primary visual cortex less than 10% of the inputs to layer 2/3 pyramidal neurons originate from layer 4 feedforward (FF) connections, whereas more than 60% of inputs come from the horizontal connections within the layer. Recently, the role of these connections has been elucidated in terms of optimizing processing for natural stimuli [27]. The efficient coding hypothesis posits that sensory neurons achieve optimal encoding by matching their tuning properties to the statistics of the natural stimuli they encode. For example, neurons in the visual (or auditory) system should be optimized for coding images (or sounds) that are representative of those found in nature. Jin et al. (2022) [27] used the statistics of natural images to derive a function that a neuron should use to compute boundary probability (**Figure 3b**). This function describes how inputs within the neuron's classical receptive field (FF) and those outside of it (horizontal) interact to modulate its response. The authors hypothesized that allocating FF inputs to distal basal dendrites and contextual inputs to their proximal parts could result in the desired input-output (I/O) integration function. Indeed, this allocation resulted in nonlinear summation of inputs due to NMDAR-dependent spatial interactions in computational models of neurons. The resulting asymmetric sigmoidal response function closely matched the boundary probability derived from natural images. This study predicts that, dendrites of pyramidal neurons provide a powerful computing substrate through which horizontal contextual connections can modulate neuronal integration function, to optimize it for natural stimuli statistics.

In addition to their spatial statistics, the temporal structure of natural stimuli also imposes some constraints on optimal neuronal processing. Neurons must be able to integrate, process, and retain information on behavioral timescales, while somatic spiking occurs on the scale of milliseconds. Neurons equipped with active dendritic conductances can overcome this discrepancy by generating long-lasting plateau potentials that extend on much longer timescales than individual somatic spikes. Leugering et al., (2023) [28] have proposed a theory of computations established by dendritic plateau potentials. According to this theory, the plateau potentials are confined to individual compartments that receive input from specific neuronal populations. The interaction of plateaus in different dendritic compartments could be used to represent sequential activations of different neuronal populations, allowing a neuron to detect complex patterns of stimuli. For example, the sequential activation of compartments receiving inputs from specific populations of place cells, can encode a more abstract notion of a path through the environment (**Figure 3c**). Representing such complex sequences with a single somatic spike results in an extremely sparse and efficient code.

**Increasing neuronal expressivity**

Expressivity refers to the ability of a neural network to represent multiple functions or mappings between inputs and outputs, which determines the range of problems that a network can solve (**Figure 3d**). Poirazi and Mel (2003) [4] demonstrated that predicting the input-output function of a single biological pyramidal neuron requires an artificial neural network (ANN) comprising at least two layers of sigmoidal subunits. Specifically, a hidden layer that corresponds to a cell's nonlinear dendrites and an output layer that represents the thresholding that occurs at the soma. These findings suggested that biological neurons, due to their active dendrites, have comparable expressivity to multilayer ANNs. Interestingly, recent experimental studies [29,30] show that certain dendritic calcium channels can enable dendrites to solve linearly non-separable problems such as the exclusive OR (XOR) problem.

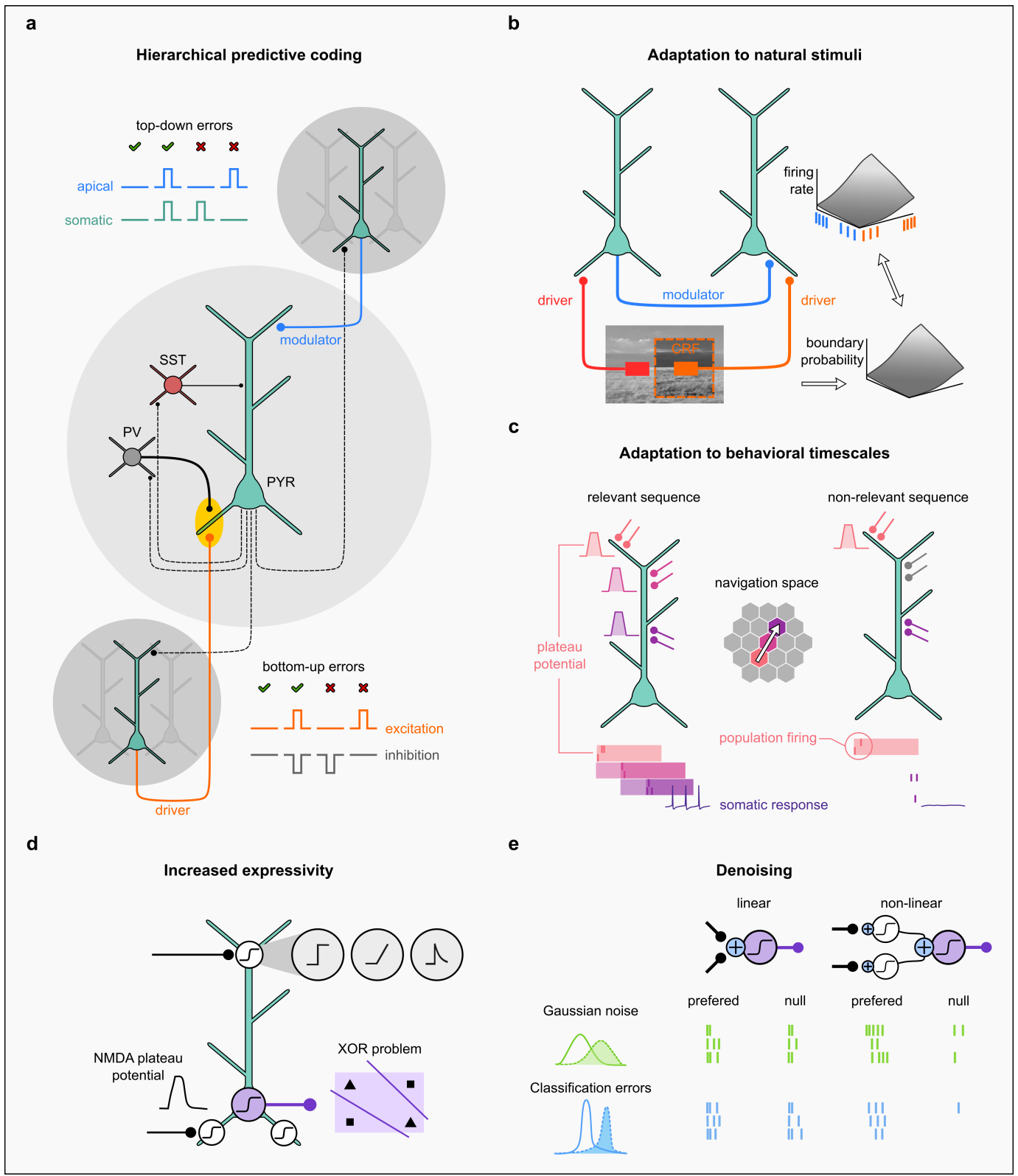

**Figure 3 | Efficient information processing in pyramidal neurons.**

**(a)** A cortical microcircuit schematic illustrates how dendrites facilitate hierarchical predictive coding [25]. Dendrites enable computation of top-down and bottom-up errors within pyramidal (PYR) neurons. Basal dendrites receive predictions from lower-level neurons, which the neuron tries to predict and cancel through lateral inhibition via parvalbumin-positive (PV) interneurons. Unpredicted inputs bypass inhibition, causing bottom-up prediction errors. Apical dendrites receive predictions from higher-level neurons that might be gated by somatostatin-positive (SST) interneurons. Mismatch between apical prediction and somatic spiking results in top-down prediction errors.

**(b)** Interaction of feedforward and horizontal inputs on basal dendrites offers an adaptation to natural visual stimuli [27]. Feedforward connections from the neuron's classical receptive field (CRF) arrive on the distal ends of basal dendrites, while horizontal connections, carrying contextual information from adjacent cells within a cortical layer, arrive on the proximal ends. The interplay of activity of proximal and distal inputs allows to recreate the boundary probability function derived from natural images through changes in the firing rate of presynaptic neurons.

**(c)** Dendritic compartments sequentially generate plateau potentials (left) in response to inputs from different neuronal populations of place cells (center) allowing a neuron to detect a complex sequence of events on behavioral timescales [28]. When one of the sequential compartments (pink) fails to evoke a plateau potential the neuron remains silent (right), thus increasing selectivity for the relevant sequence.

**(d)** Dendrites equipped with active conductances, that enable generating local plateau potentials, increase expressivity of individual neurons, allowing them to solve linearly non-separable problems such as the exclusive OR (XOR) problem.

**(e)** Nonlinear dendrites, in contrast to linear ones, provide effective strategies for mitigating neuronal noise. While both linear and nonlinear dendrites can reduce Gaussian noise, nonlinear dendrites excel in decreasing (mis)-classification errors [33], a common type of noise in the brain. These errors occur when neurons spike in response to non-preferred (null) stimuli or remain silent when exposed to preferred ones.

---

These action potentials exhibit maximum amplitude for stimuli that are at threshold-level, but are dampened for stronger stimuli. This discovery is significant as it expands the range of computations that can be performed by individual neurons, without the need for multi-layered networks which were thought to be necessary for such tasks.

While earlier studies used artificial neural networks (ANNs) to replicate the time-averaged firing rate of realistic neuron models, a recent study by Beniaguev et al. (2021) [31] took a different approach. They implemented a temporally convolutional deep neural network (DNN) that could also capture the temporal aspect of neuronal activity with millisecond precision. The study demonstrated that a network's depth (number of its layers) can serve as a proxy for assessing a neuron's expressivity. Specifically, to capture the input-output function of a realistic model of layer 5 pyramidal neuron (L5PN), it required a DNN consisting of five to eight layers. Removing NMDA receptors substantially decreased the neuron's complexity, as its output could be predicted by a single hidden layer. This study highlights that a neuron's expressivity is directly dependent on the presence of active dendritic mechanisms.

Wybo et al. (2022 arXiv) [32] further supported this idea by showing that NMDA spikes could modulate the I/O relationship in an abstract L5PN model receiving perisomatic FF inputs modulated by FB dendritic inputs. The number of dendritic compartments with NMDA spikes determined the slope and threshold of a rectified linear unit (ReLU) function, representing the neuron's I/O relationship. Remarkably, in this study NMDA modulation allowed for learning multiple tasks by modifying only FB weights, while keeping the previously learned FF weights untouched. These findings suggest that a highly expressive neuron can adapt to different contexts by switching between different I/O regimes. On a network level, increased expressivity of individual neurons can promote sparse coding, which reduces the number of neurons and overall computational cost required for a given task.

**Mitigating the negative effects of neuronal noise**

Variability is an inherent property of neuronal activity. It can arise from numerous sources, both external environmental factors during sensory transduction, and intrinsic properties of the nervous system. The latter include the stochasticity of biochemical processes such as channel gating, vesicle release, and neurotransmitter diffusion. As the main purpose of neuronal activity is to communicate useful information, these random disturbances can be thought of as "noise" that obscures the relevant signal. Noise can interfere with the precision and reliability of neural communication, leading to a loss of information and affecting the efficiency of neuronal coding.

Previous studies have focused on linearly integrating neurons, assuming that the primary mechanism for denoising is the averaging of a large number of synaptic inputs. However, these studies did not take into account the presence of active ionic conductances in dendrites that can generate dendritic spikes. Poleg-Polsky (2019) [33] demonstrated that active nonlinear dendrites, in contrast to linear ones, are more efficient in the presence of misclassification errors (**Figure 3e**). This type of noise — when a neuron spikes in response to a non-preferred stimulus or remains silent when exposed to a preferred one — is ubiquitous in the brain and resistant to averaging. The additional thresholding step introduced by dendritic spike generation, allows neurons to mitigate it more effectively. In particular, neurons with active dendrites outperform linearly integrating neurons in a directional discrimination task in the presence of directional misclassification errors. Thus, dendritic spikes challenge noise tolerance rules derived for linear neurons.

Although noise is typically assumed to degrade performance, it can in fact have a positive effect on information processing. A common example is the amplification of weak signals through stochastic resonance, which occurs in the presence of a certain level of noise. Specifically, weak signals that ride on top of noisy background activity can more easily cross the threshold for dendritic spiking, thus propagating to the soma. This mechanism enhances information transmission between neurons, allowing them to communicate more information per action potential. As a result, it reduces the metabolic cost and increases energy efficiency in coding [34]. Noise could also play an important role in the context of hierarchical predictive coding. Specifically, noise can relax the constraints on the delay in inhibition, when balancing bottom-up inputs and allow for neural sampling [25]. Finally, noise can produce new dynamics such as a bistability in I/O transformation due to slow calcium activation — a phenomenon absent in a noiseless condition [35].

## Information storage

### Optimizing learning and storage capacity

In addition to efficient processing, dendrites enhance learning and memory storage. Poirazi and Mel (2001) [36] demonstrated that nonlinear integration in active dendrites allows a model neuron to distinguish more input configurations distributed across its branches, providing a distinct advantage over linear summation. Moreover, through structural plasticity, neurons have access to an additional high-capacity storage reservoir, namely the arrangement of synaptic connections, instead of relying solely on changes in synaptic strength. As a result, the combination of nonlinear dendrites and structural plasticity increases the storage capacity of a model neuron by an order of magnitude. Similar benefits were observed in hippocampal network models. When fast spiking interneurons were equipped with nonlinear dendritic integration, the encoding of a single memory required significantly fewer neurons compared to the linear case, thus providing important resource savings [37] (**Figure 4a**). Moreover, nonlinear dendritic integration of separate information streams increased the capacity to store and retrieve memory engrams in two-compartment models of hippocampal pyramidal cells [38]. Notably, the interference of memories, which can lead to the loss of previously stored memories, was minimal even when a large number of highly similar memories were stored.

While memory interference can lead to confusion and hinder survival, binding of information is critical for the formation of associative and episodic memories. Thus, brains must optimize the tradeoff between memory binding and interference. Synaptic clustering is a potential solution to this problem (**Figure 4b**). Numerous studies have shown that synaptic clustering can emerge from the co-activation of nearby inputs, due to their efficiency in activating local non-linearities and facilitating cooperative plasticity (for a review see: Kastellakis et al., 2015; 2019 [39,40]). Such clustering binds together separate

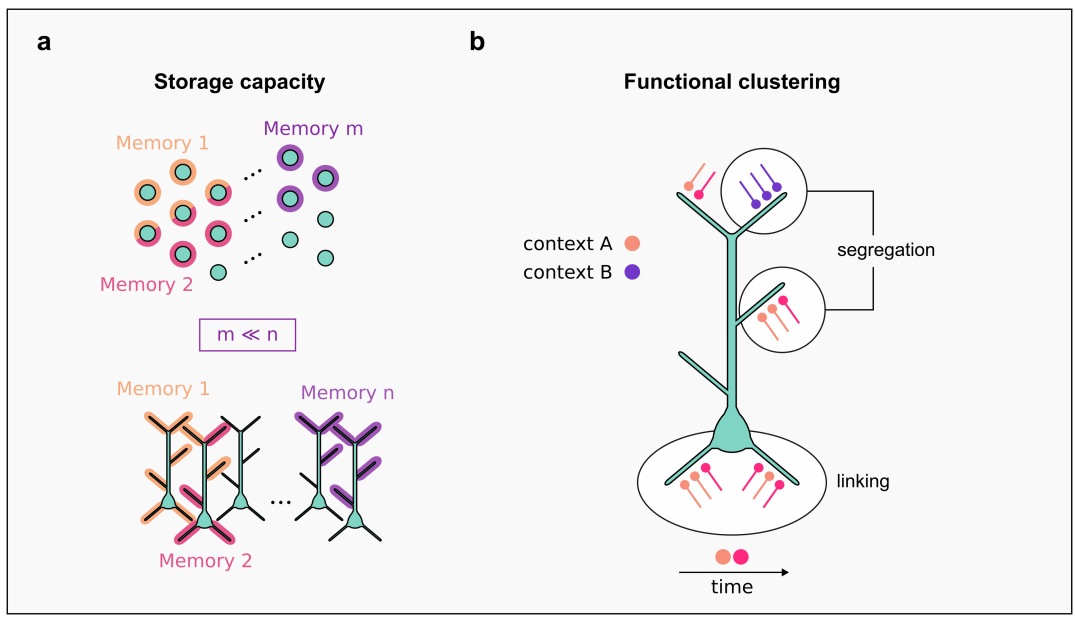

**Figure 4 | Efficient storage of information in pyramidal neurons.**

**(a)** Storage capacity in a population of point neurons - equivalent to linear dendritic integration - (top, memories) is much lower than storage capacity in a population with nonlinear dendritic integration (bottom, memories) given that the number of neurons in both populations is the same.

**(b)** Clustering of synaptic inputs based on their functional similarity can serve either to bind similar memories via co-clustering within a branch (e.g. linking across time [42]) or to segregate different memories by distributing memory-specific clusters to different branches, thus preventing memory interference (e.g. [44]).

---

information streams, allowing the formation of associative memories [41,42] and can link different memories separated by hours [42,43]. Interference, in this case, can be minimized by allocating functionally different inputs to separate branches [41,44].

Finally, the ability to rewire through structural plasticity has been associated with faster learning. CCR5 knock-out mice, which exhibit higher than normal spine turnover dynamics, learn a fear conditioning task much faster than controls. Faster learning is likely due to an increase in the formation and stability of synaptic clusters, which is associated with sparser encoding. Stable spine clusters also serve as a means for protecting memories from subsequent modifications through stochastic rewiring [45]. Synaptic turnover was also shown to increase the efficiency of learning in bio-inspired spiking neural networks (SNNs) applied to machine learning tasks [46]. Specifically, learning to discriminate between two classes of MNIST digits was faster and achieved using significantly fewer synaptic weights in the presence of synaptic turnover compared to a model without synaptic turnover. This study provided exciting new findings on the means by which synaptic turnover can improve the efficiency of learning. Overall, the combination of nonlinear dendrites and structural plasticity can optimize learning, storage, and discrimination of memories through the formation of stable synaptic clusters. Given the space limitations of the neuronal substrate, storing more memories in the same population of neurons, and achieving this storage via the use of fewer synapses, is significantly more resource-efficient.

## Conclusion

Throughout nature, simple laws give rise to emergent complex behaviors, from protein folding to the organization of ecosystems. These laws often revolve around the principles of efficiency and energy saving. The brain offers a particularly compelling example of such a complex system, thus studying efficiency in the brain might be crucial for understanding it. Moreover, by emphasizing the importance of efficiency, research in neuroscience shifts the focus from asking "how" questions to inquiring "why" the brain works the way it does, ultimately fostering the development of a more comprehensive theoretical framework.

The study of efficiency in the brain has a long history, with Barlow's pioneering work from half a century ago proposing that neurons communicate as much information as possible through as few spikes as possible [47]. However, while revealing important principles of efficient coding, early works did not take into account the elaborate dendritic trees that neurons possess. Over the last few decades, computational and experimental studies have demonstrated that dendrites do not merely convey synaptic inputs to the soma, but perform complex computations due to their nonlinear properties. In this review, we presented multiple pieces of evidence that nonlinear dendritic computations increase the efficiency of information processing and storage in cortical and hippocampal pyramidal neurons. While dendrites may not be strictly necessary for abstract computations [48], their evolutionary advantage becomes clear in real-world scenarios when resources such as energy and space are limited.

These ideas not only expand our understanding of biological networks, but also have far-reaching implications for the fields of artificial intelligence and neuromorphic computing. According to recent research, incorporating dendritic computational principles into artificial neural networks [49] or neuromorphic systems [50,51] substantially enhances their efficiency compared to classical-architecture-based systems. What lies ahead is an exciting interplay between neuroscience, machine learning and hardware experts, so as to capitalize on the wisdom of the brain and fully exploit the properties of dendrites to advance the efficiency of artificial systems.

# Conflict of interest

The authors declare no conflict of interest.

# Acknowledgements

We would like to thank Dr. Spyridon Chavlis and other members of the Poirazi lab for their valuable feedback on the manuscript. This work was supported by NIH (1R01MH124867-02), the European Commission (H2020-FETOPEN-2018-2019-2020-01, NEUREKA GA-863245 and H2020 MSCA ITN Project SmartNets GA-860949), and the Einstein Foundation Berlin, Germany (visiting fellowship to PP, EVF-2019-508).

# References and recommended reading